\documentclass[twocolumn,aps,10pt,showkeys,showpacs,preprintnumbers,prd,superscriptaddress,nofootinbib]{revtex4-1}
\usepackage{graphicx}	
\usepackage{amssymb}
\usepackage{dcolumn}
\usepackage{mathtools}
\usepackage{amsmath}
\usepackage{xcolor}
\usepackage{color}
\usepackage[normalem]{ulem}
\usepackage{theorem}
\usepackage{subfigure, rotating, bm, array}
\usepackage[pagebackref=false, colorlinks=true]{hyperref}
\hypersetup{linkcolor=blue, citecolor=blue,urlcolor=blue}

\begin{document}
\title{Quasinormal Modes and Stability Analysis of the JMN-1 Naked Singularity}

\author{Akshat Pathrikar}
\email{akshatpathrikar014@gmail.com}
\affiliation{International Centre for Space and Cosmology, Ahmedabad University, Ahmedabad 380009, Gujarat, India}
\author{Parth Bambhaniya}
\email{grcollapse@gmail.com}
\affiliation{Instituto de Astronomia, Geofísica e Ciências Atmosféricas, Universidade de São Paulo, IAG, Rua do Matão 1225, CEP: 05508-090 São Paulo - SP - Brazil.}
\author{Pankaj S. Joshi}
\email{psjcosmos@gmail.com}
\affiliation{International Centre for Space and Cosmology, Ahmedabad University, Ahmedabad 380009, Gujarat, India}
\author{Elisabete M. de Gouveia Dal Pino}
\email{dalpino@iag.usp.br}
\affiliation{Instituto de Astronomia, Geofísica e Ciências Atmosféricas, Universidade de São Paulo, IAG, Rua do Matão 1225, CEP: 05508-090 São Paulo - SP - Brazil.}
\date{\today}
\begin{abstract}
In this paper, we perform a comprehensive analysis of the quasinormal modes in an external geometry of the Joshi-Malafarina-Narayan (JMN-1) naked singularity by investigating its response to linear perturbations, including gravitational and electromagnetic perturbations. To analyze the stability of the JMN-1 naked singularity under axial perturbations, 
we compute the quasinormal mode frequencies using the Wentzel-Kramers-Brillouin method. The quasinormal mode frequencies provides information about the stability of spacetime, with the real part of the frequency determining the oscillation rate and the imaginary part governing the decay or growth of perturbations. Our results indicate that by imposing appropriate boundary conditions, we find that the background spacetime of JMN-1 naked singularity remains dynamically stable under axial perturbations.

\vspace{.6cm}
$\boldsymbol{Key words}$ : Quasinormal modes, Stability, Naked singularities, WKB method.
\end{abstract}
\maketitle

\section{Introduction}

The foundational model that describes the gravitational collapse of a spherically symmetric and homogeneous dust cloud was introduced by Oppenheimer, Snyder, and Dutt (OSD) \cite{Oppenheimer:1939ue}. Their findings suggest that such a collapse ultimately results in the formation of a black hole. In this framework, trapped surfaces emerge around the central region prior to the occurrence of a central spacelike singularity. As a result, the singularity is causally isolated from the rest of spacetime, leading to the presence of an event horizon. This singularity is not just a point or region within spacetime, but rather represents a fundamental limit to the structure of spacetime itself. In other words, it represents the boundary of the spacetime manifold where all the non-spacelike geodesics are incomplete and the volume element of the Jacobi field will get vanished infinitely.  However, the OSD model is based on idealized assumptions, including a uniform density distribution and the absence of pressure within the collapsing stellar body, making it a highly simplified representation of gravitational collapse.  

Extending the OSD framework, Roger Penrose proposed the Cosmic Censorship Conjecture (CCC), which postulates that the final singularity resulting from gravitational collapse must always remain concealed within an event horizon, ensuring that the collapse invariably leads to black hole formation \cite{penrose}. Despite extensive efforts to establish a rigorous mathematical formulation or proof for CCC, no conclusive resolution has been reached. This remains one of the most fundamental open problems in gravitational physics, with significant implications for black hole theory and astrophysical applications. Consequently, various studies have explored more physically realistic collapse scenarios within Einstein gravity, incorporating inhomogeneous matter distributions and nonzero pressure profiles \cite{joshi,Joshi:2011zm,Joshi:2011rlc,Joshi:2024gog,goswami,Deshingkar:1998ge,Jhingan:2014gpa,Joshi:2013dva}. These studies raise the important question of how to observationally distinguish naked singularities from conventional black hole spacetimes. Therefore, various observational characteristics of naked singularities have been investigated, including their shadows and accretion disk properties \cite{Joshi:2013dva,Saurabh:2023otl,gyulchev,Tahelyani:2022uxw,Kovacs:2010xm,Guo:2020tgv,Joshi:2020tlq,Saurabh:2022jjv,Solanki:2021mkt,Bambhaniya:2021ybs,Bambhaniya:2024lsc,Vagnozzi:2022moj, Bambi:2019tjh}, periastron precession of relativistic orbits \cite{Bambhaniya:2019pbr,Joshi:2019rdo,Dey:2019fpv,Bam2020,Bambhaniya:2025xmu,Bambhaniya:2022xbz,Bambhaniya:2021jum}, pulsar timings \cite{Kalsariya:2024qyp}, tidal force effects \cite{Madan:2022spd}, energy extraction mechanisms \cite{Patel:2023efv,Patel:2022jbk,Acharya:2023vlv}, etc.

The work in \cite{Joshi:2011zm}, extensively examines the equilibrium configurations of collapsing clouds influenced by gravitational forces under the assumption of zero radial pressure but nonzero tangential pressures, leading to an anisotropic fluid model. Under such conditions, the collapse of a massive matter cloud can give rise to a JMN-1 naked singularity. Like several other solutions in general relativity, the Joshi-Malafarina-Narayan (JMN-1) spacetime does not emerge from a globally hyperbolic Cauchy development \cite{Joshi:2024gog}. While traditional determinism implies the existence of a global Cauchy surface, many physically significant spacetimes, including Kerr, Reissner-Nordström, and Kerr-Newman, do not satisfy this requirement and may even exhibit causality violations \cite{Joshi:2024gog}. However, unlike these cases, the JMN-1 spacetime does not contain closed timelike curves, ensuring that causality is preserved. Studies indicate that under reasonable energy conditions and regular initial data, naked singularities can naturally arise from gravitational collapse \cite{joshi,Joshi:2011zm,Joshi:2011rlc}. Given that the JMN-1 spacetime maintains causal consistency, it represents a physically valid alternative to black holes in gravitational collapse scenarios.  

Recent discoveries by the Event Horizon Telescope (EHT) collaboration have greatly contributed to the study of black hole imaging and observations, enhancing our understanding of these extreme astrophysical compact objects. Their findings indicate that the possibility of Sgr A* being a JMN-1 naked singularity cannot be ruled out based on the metric test \cite{EventHorizonTelescope:2022xqj}. Also, recent study demonstrated that many naked singularity models feature inner turning points for timelike and lightlike geodesics (non-spacelike), which leads to the formation of an accretion-powered photosphere within the shadow region \cite{Broderick:2024vjp}. This implies that accretion shocks should occur inside the photon sphere. However, observations of Sgr A* and M87* by the EHT collaboration suggest that the accretion flow remains coherent up to the photon sphere radius. As a result, most naked singularity models can be, in principle, excluded, except JMN-1 and Janis-Newman-Winicour (JNW), both are of type P0j\footnote{The P0j-type singularities are defined by a finite angular momentum, where timelike geodesics can reach the singularity. This corresponds to the parameter range in which an unstable photon orbit exists \cite{Broderick:2024vjp}.}.
These exceptions lack the characteristic inner turning points for non-spacelike geodesics before reaching the singularity, making it challenging to detect accretion-driven shocks or photosphere within their shadow.

However, a crucial aspect of any spacetime solution in general relativity is its dynamical stability under small perturbations. Stability analysis provides insight into whether a given solution represents an equilibrium configuration or if small perturbations can lead to significant deviations, thereby altering the physical interpretation of the solution. One of the most effective methods for investigating stability in black hole and naked singularity spacetimes is the study of quasinormal modes (QNMs). These are the characteristic oscillations of a perturbed spacetime, governed by complex frequencies that encode information about the stability and response of the spacetime to external disturbances. The real part of the QNM frequency determines the oscillation frequency, while the imaginary part dictates the decay or growth of the perturbation. 

The study of QNMs in black hole spacetimes has been extensively explored \cite{Konoplya:2019,Bhattacharya:2023,Konoplya:2003,Matyjasek_2017,Konoplya:2023,Zhao:2024,Jusufi:2020,Abdalla:2006, Ghosh:2023etd}, revealing fundamental aspects of black hole dynamics, the nature of event horizons, and even potential observational signatures in gravitational wave astronomy \cite{Kokkotas:1999bd,Berti:2009kk}. However, the investigation of QNMs in naked singularity spacetimes remains an active and relatively less explored area of research. As we have pointed out above, based on metric tests and shadow images, the JMN-1 and JNW naked singularities are the most promising candidates to mimic conventional black hole models \cite{EventHorizonTelescope:2022xqj,Saurabh:2023otl,gyulchev,Vagnozzi:2022moj}. The QNMs and stability of the JNW naked singularity have been explored in \cite{Chowdhury:2020,Stashko:2023ffs}. On the other hand, the JMN-1 spacetime, as a well-defined naked singularity solution, presents an excellent opportunity for studying the behavior of perturbations in such a background geometry. Whether it exhibits stable oscillatory behavior or instability remains an important aspect to investigate.

The stability analysis via QNMs can reveal whether the JMN-1 solution is physically viable or if perturbations lead to growing modes frequencies that may ruled out such spacetimes. In this work, we analyze QNM frequencies using the Wentzel-Kramers-Brillouin (WKB) method in the background metric of JMN-1 naked singularity by considering linear perturbation expansion of the metric and investigate the stability of this solution under gravitational and electromagnetic perturbations. In the stability analysis based on test fields, “scalar” refers to an 
external Klein–Gordon field propagating on a fixed background, treated analogously to electromagnetic or gravitational test fields. Such scalar field perturbations, by definition, do not appear in the axial (odd-parity) sector; instead, they are associated with the polar (even–parity) sector.

We have the following arrangement of the paper. In section (\ref{sec:2}), we perform the analysis of perturbations in an external spacetime of JMN-1 naked singularity. In section (\ref{sec:3}), we discuss the boundary conditions for the JMN-1 naked singularity spacetime. In section (\ref{sec:4}), we carry out the analysis of WKB method to study the perturbations. In section (\ref{sec:resullts}), we study the frequencies of fundamental QNM. In section (\ref{sec:5}), we discuss and conclude the results of this work. Throughout the paper, we use the metric sign as $-,+,+,+$ and the units $G = c = 1$.

\section{Perturbations in JMN-1 Spacetime}
\label{sec:2}
The JMN-1 spacetime is supported by an anisotropic fluid, where the stress-energy tensor exhibits varying pressure components in different directions. Specifically, the energy density $\rho$ and pressure components $p$ are given by \cite{Joshi:2011zm}:  
\begin{equation}  
    \rho=\frac{M_0}{r^2}, \; \; \;  
    p_r=0, \; \; \;  
    p_{\theta}=p_{\phi}=\frac{M_0}{4(1-M_0)}\rho.  
\end{equation}  
This configuration implies that while the radial pressure is absent, the tangential pressure plays a crucial role in preventing the formation of trapped surfaces, ultimately allowing for the emergence of a naked singularity. Importantly, the stress-energy tensor adheres to standard energy conditions, indicating that the matter distribution is physically viable. The corresponding metric for the JMN-1 naked singularity spacetime is expressed as:  
\begin{equation}  
     ds^2 = - (1- M_{0}) \left( \frac{r}{R_b}\right)^{\frac{M_{0}}{1 - M_{0}}} dt^2 + \frac{1}{(1- M_{0})}  dr^2 +\, r^2  d\Omega^2,  
     \label{JNWst}  
\end{equation}  
where $d\Omega^2= d\theta^2+\,sin^2\,\theta\,d\phi^2$. Here, $R_b$ denotes the boundary radius at which the interior JMN-1 spacetime seamlessly transitions into the exterior Schwarzschild spacetime, defining the extent of the matter distribution. 

The parameter $M_0$ is a dimensionless quantity that characterizes the compactness of the object and the intensity of anisotropic pressure effects. It must satisfy the constraint $0<M_0<4/5$ to ensure that the speed of sound remains below the speed of light. The effective sound speed of the equilibrium state, given by $c_s^2= p/\rho$, must remain subluminal, thus necessitating $M_0 < 4/5$ \cite{Joshi:2011zm}. The Schwarzschild mass of the system is given by $M=\frac{1}{2} M_0 R_b$, establishing a direct relationship between $M_0$ and the total gravitational mass. Physically, a higher $M_0$ results in stronger gravitational effects near the core, yielding a more compact structure. Notably, $R_b$ must not be less than $2.5 M$ to satisfy the sound speed condition.

The metric describing this spacetime is constructed based on a compact high-density region in a vacuum, which requires that the overall spacetime configuration be asymptotically flat. To achieve this, the JMN-1 spacetime is matched smoothly to an exterior Schwarzschild spacetime at the boundary radius $r=R_{b}$. The matching conditions between these two spacetimes require that $(i)$ the extrinsic curvatures ($K_{ab}$) of the internal and external spacetimes should be matched at the hypersurface, where the extrinsic curvatures are expressed in terms of the covariant derivative of normal vectors on the hypersurface:
    \begin{equation}
K_{ab}=e^{\alpha}_ae^{\beta}_{b}\nabla_{\alpha}\eta_{\beta},
    \end{equation}
    where $e^{\alpha}_a$ and 
    $e^{\beta}_b$ 
    are the tangent vectors on the hypersurface and 
    $\eta_{\beta}$
    is the normal to that hypersurface. Also, $(ii)$ the induced metrics of the exterior and interior geometries coincide at the junction surface. Since the JMN-1 naked singularity spacetime is characterized by zero radial pressure, the extrinsic curvatures of the interior JMN-1 region and the exterior Schwarzschild region align smoothly at $r=R_{b}$ \cite{Bambhaniya:2019pbr}. 
One can smoothly match this interior JMN-1 spacetime to the exterior Schwarzschild spacetime at $r = R_b$ and the line element can be expressed as:
 \begin{equation}
    ds^2 = - \left(1 - \frac{M_0 R_b}{r} \right) dt^2 + \left(1 - \frac{M_0 R_b}{r} \right)^{-1} dr^2 + r^2 d\Omega^2.
    \label{matchedjmn}
\end{equation}
It is important to emphasize that throughout this paper, our analysis is restricted to the exterior region of the JMN-1 spacetime, which ensures asymptotic flatness \cite{Bambhaniya:2019pbr}. In particular, we consider this scenario to be analogous to the Schwarzschild black hole, where stability analyses are typically conducted in the external region extending from $2M$ to spatial infinity. In the black hole case, the ingoing plane waves are entirely absorbed by the event horizon located at the inner boundary. In contrast, for the JMN-1 naked singularity, perturbations may be partially scattered, reflected, or fully absorbed at the inner boundary or at the singularity. The behavior of these ingoing waves might depend on the specific matter profile. In this work, we focus solely on the external geometry of the JMN-1 naked singularity to avoid additional complexities. The study of perturbations within the boundary radius $R_b$ in the JMN-1 spacetime will be addressed separately in future work, as it requires alternative methods to investigate the stability of the interior region.    

The general form of a static, spherically symmetric spacetime can be recast into the line element: 

\begin{equation}
ds^2 = -f(r) \, dt^2 + \frac{dr^2}{f(r)} + r^2 d\Omega^2 .
\end{equation}
We now introduce small perturbations \( h_{\mu\nu} \) to the background metric \( \bar{g}_{\mu\nu} \) such that the perturbed metric \( g_{\mu\nu} \) becomes:

\begin{equation}
    g_{\mu\nu} = \bar{g}_{\mu\nu} + h_{\mu\nu}, \quad \text{where} \quad \frac{|h_{\mu\nu}|}{|\bar{g}_{\mu\nu}|} \ll 1.
\end{equation}
In addition to gravitational perturbations, we consider electromagnetic field perturbations on the fixed background spacetime. For this field, we work with the vector potential $A_\mu$, which is decomposed as
\begin{equation}
    A_\mu = \bar{A}_\mu + \delta A_\mu,
\end{equation}
where $\bar{A}_\mu$ is the background potential (vanishing in the case of an uncharged black hole spacetime), and $\delta A_\mu$ encodes the perturbations.

Now for the gravitational case, the perturbed metric gives rise to the perturbed Christoffel symbols:

\begin{equation}
    \Gamma^\alpha_{\mu\nu} = \bar{\Gamma}^\alpha_{\mu\nu} + \delta\Gamma^\alpha_{\mu\nu},
\end{equation}
where \( \bar{\Gamma}^\alpha_{\mu\nu} \) are the Christoffel symbols of the unperturbed metric, and

\begin{equation}
    \delta\Gamma^\alpha_{\mu\nu} = \frac{1}{2} \bar{g}^{\alpha\beta} \left( \nabla_\mu h_{\nu\beta} + \nabla_\nu h_{\mu\beta} - \nabla_\beta h_{\mu\nu} \right).
\end{equation}
This results in the perturbed Ricci tensor:

\begin{equation}
    R_{\mu\nu} = \bar{R}_{\mu\nu} + \delta R_{\mu\nu},
\end{equation}
where,
\begin{equation}
    \delta R_{\mu\nu} = \nabla_\nu \delta\Gamma^\alpha_{\mu\alpha} - \nabla_\alpha \delta\Gamma^\alpha_{\mu\nu}.
    \label{pert_Einst}
\end{equation}
Here, \( \nabla_\mu \) is the covariant derivative with respect to the background metric \( \bar{g}_{\mu\nu} \).

Due to the spherical symmetry of the background spacetime, we can decompose the perturbations into odd and even types. 

In this work, we focus exclusively on the axial (odd-parity) perturbations of the background metric. We assume that the anisotropic matter fluid constituting the JMN-1 naked singularity within $R_b$ does not contribute to these perturbations. In many scenarios, anisotropic fluids with purely tangential pressure do not affect odd-parity perturbations \cite{Chowdhury:2020,C:2024cnk,Zhang:2024bov,Blazquez-Salcedo:2013ava,Kobayashi:2012kh}. This is because odd-parity perturbations involve shear-like distortions of the metric rather than scalar quantities such as energy density or pressure. In such cases, the axial perturbations obey a Regge-Wheeler (RW)-like equation. However, certain specific models of anisotropic fluids can contribute to axial perturbations, particularly when shear viscosity is present or when the stress-energy tensor exhibits nonzero off-diagonal components in an orthonormal frame.
Since we consider JMN-1 as a spherical symmetric spacetime, the evolution of the axial perturbation for the asymptotically flat spacetime is governed by the following field equation:

\begin{equation}
    \delta R_{\mu\nu} = 0.
\end{equation} 
The perturbation variables $h_{\mu\nu}$ can be expanded in a series of spherical harmonics and can be separated into axial and polar parts\footnote{In this case ``axial'' and ``polar'' \cite{Chandrasekhar:1985kt} corresponds to ``odd'' and ``even'' parities according to the initial paper \cite{PhysRev.108.1063}.}, which can be treated independently. In this paper, we  focus solely on the axial perturbations. The matrix of the axial gravitational perturbations $h_{\mu\nu}$ in the Regge-Wheeler gauge take the following form,

\begin{equation}
    h_{\mu\nu}^{\text{axial}} = 
    \begin{bmatrix}
        0 & 0 & 0 & h_0(t,r) \\
        0 & 0 & 0 & h_1(t,r) \\
        0 & 0 & 0 & 0 \\
        h_0(t,r) & h_1(t,r) & 0 & 0
    \end{bmatrix}
    \left( \sin\theta \frac{\partial}{\partial \theta} \right) P_l (\cos \theta).
    \label{pert_axial}
\end{equation}
where, $h_0(t,r)$ and $h_1(t,r)$ are two unknown functions, and $P_l(x)$  is the Legendre polynomial\footnote{Due to the spherical symmetry of the background configuration, more general case that involves $Y_{lm}$, leads to the same master equation and the same QNM frequencies. The perturbations with $l=1$ can be removed by gauge transformation \cite{Vishveshwara:1970}. } with $l\geq2$. After substitution of Eqn.(\ref{pert_axial}) to the perturbed Einstein equations Eqn.(\ref{pert_Einst}) and retaining only linear terms, we obtain:

\begin{align}
\delta R_{t\phi} &= - \frac{\partial}{\partial r} \left[ \frac{f(r)}{2} \left( \frac{\partial h_1}{\partial t} - \frac{\partial h_0}{\partial t} \right) (\sin\theta \, \partial_{\theta}) P_l(\cos\theta) \right] \nonumber \\ 
&\quad + \frac{\partial}{\partial \theta} \left[ \frac{h_0}{2r^2} \left( \frac{\partial}{\partial \theta} (\sin\theta \, \partial_{\theta}) P_l(\cos\theta) \right) \right] = 0,
\end{align}

\begin{align}
\delta R_{r\phi} &= - \frac{\partial}{\partial t} \left[ -\frac{1}{2} f(r) \left( \frac{\partial h_0}{\partial r} - \frac{\partial h_1}{\partial t} \right) (\sin\theta \, \partial_{\theta}) P_l(\cos\theta) \right] \nonumber \\ 
&\quad + \frac{\partial}{\partial \theta} \left[ -\frac{h_1}{2r^2} \left( \frac{\partial}{\partial \theta} (\sin\theta \, \partial_{\theta}) P_l(\cos\theta) \right) \right] = 0,
\end{align}

\begin{align}
\delta R_{\theta\phi} &= - \frac{\partial}{\partial t} \left[ \left(-\frac{h_0}{2 f(r)} \right) \frac{\partial}{\partial \theta} \left( \sin\theta \, \frac{\partial}{\partial \theta} P_l(\cos\theta) \right) \right] \nonumber \\
&\quad - \frac{\partial}{\partial r} \left[ \frac{f(r)}{2} h_1 \frac{\partial}{\partial \theta} \left( \sin\theta \, \frac{\partial}{\partial \theta} P_l(\cos\theta) \right) \right] = 0
\end{align}

Defining 
\begin{equation}
\psi = \left( 1 - \frac{M_0 R_b}{r} \right) \frac{h_1}{r}
\end{equation}
This substitution redefines the perturbation variable $\psi$ to factor out the background metric's influence on the perturbation function $\frac{h_1}{r}$ so that we get the RW equation into a more manageable form.\\

In the JMN-1 metric Eqn.(\ref{matchedjmn}) sets $ds^{2} = 0$  
and the tortoise coordinate is defined as follows;
\begin{equation}
    dr^* = (1 - M_0R_b/r)^{-1} dr,
\end{equation}
recollecting the substitutions and the equations above we get a Schr\"odinger wave like equation, also known as the RW equation, which corresponds to the wave equation for the perturbations,

\begin{equation}
\frac{d^2 \psi}{dt^2} - \frac{d^2 \psi}{dr_*^2} + V(r) \psi = 0,
\label{RW}
\end{equation}
where the RW potential is given by 
\begin{equation}
V_{\text{eff}}(r) = \left( 1 - \frac{M_0 R_b}{r} \right) \left[ \frac{l(l+2)}{r^2} - \frac{3 M_0 R_b}{r^3} \right].
\label{potential}
\end{equation}
Similarly, the general relativistic wave equations for the electromagnetic ($A_\mu$) field can be written as follows:
\begin{align}
    \frac{1}{\sqrt{-g}} \, \partial_\mu \left( F_{\rho\sigma} \, g^{\rho\nu} g^{\sigma\mu} \sqrt{-g} \right) &= 0, \label{eq:em_field}
\end{align}
where $F_{\mu\nu} = \partial_\mu A_\nu - \partial_\nu A_\mu$ is the electromagnetic field strength tensor.
The effective potential for the electromagnetic ($s=1$) fields then takes the form
\begin{equation}
    V_{\text{eff}}(r) = f(r) \left[ \frac{\ell(\ell+1)}{r^2}  \right],
    \label{eq:potential}
\end{equation}
where $\ell = s, s+1, s+2, \dots$ are the multipole numbers.\\

The RW potential \( V_{\text{eff}}(r) \) plays an essential role in analyzing the stability of spacetime under metric perturbations. The shape of this potential determines whether perturbations decay over time or grow unbounded. By solving the RW equation, which contains the potential term, we can extract the quasibound states. Since the RW potential for the JMN-1 spacetime happens to be independent of time, we can convert Eqn.(\ref{RW}) to Eqn.(\ref{schrodinger}) by decomposing $\psi(r_*, t)$  into a spatial  and a temporal part as $\psi(r_*, t) = \tilde{\psi}(r_*) e^{-i\omega t}$. This transformation introduces the frequency term $\omega$, allowing us to study the QNMs by solving a Schrödinger-like equation;
\begin{equation}
\frac{d^2 \tilde{\psi} (r_*)}{dr_*^2} + \left( \omega^2 - V_{eff}(r_*) \right) \tilde{\psi} (r_*) = 0.
\label{schrodinger}
\end{equation}
This master equation for QNMs governs the perturbations in a given spacetime and regulates the behavior of wave propagation in a curved background. By solving this equation, one can determine the stability aspects of any given spacetime. The extracted QNM frequencies, \( \omega \), characterize the damping and oscillatory nature of these perturbations.

\section{Boundary Conditions}
\label{sec:3}
In order to extract QNM frequencies from the compact objects, certain boundary conditions are needed to be imposed for the master equation. For black hole spacetimes, the standard approach is to impose a purely outgoing wave condition at spatial infinity, reflecting the fact that QNMs represent perturbations that radiate outward without being reflected back into the system. However, since numerical computations are performed on a finite domain, it is necessary to approximate spatial infinity by selecting a sufficiently large but finite radial coordinate where the outgoing wave behavior is well established.

In our analysis of the JMN-1 naked singularity, we impose the outer boundary condition at $r = 100M$, which effectively serves as an approximation for spatial infinity. This choice is motivated by the asymptotic behavior of gravitational perturbations in nearly flat spacetimes, where the wave function behaves as, 

\begin{equation}
    \Psi(r) \sim e^{-i\omega(t - r_*)}
\end{equation}
 At sufficiently large $r$, the perturbation equation stabilizes, ensuring that the outgoing wave condition is accurate. Studies of QNMs in Schwarzschild and Kerr spacetimes \cite{Kokkotas:1999bd, Berti:2009kk} have demonstrated that setting the outer boundary at moderate distances is sufficient to extract accurate mode frequencies, as further increases in $r$ do not significantly affect the results. 

Unlike black hole spacetimes, where the inner boundary for QNM analysis is set at the event horizon with a purely ingoing wave condition, the JMN-1 naked singularity presents a unique challenge due to the absence of an event horizon. Since the central singularity at $r = 0$ is globally naked, we must introduce an alternative inner boundary condition to ensure a well-posed perturbation problem. Therefore, we impose an inner boundary at $r= R_b$. Classically, it is expected that at the inner boundary, ingoing waves are fully absorbed by the central singularity. As a result, the wave function near this region takes the form
\begin{equation}
    \Psi(r) \sim e^{-i\omega(t + r_*)},
\end{equation}
representing purely ingoing behavior. We take $M_0 = 0.7$ and $R_b = 2.8571$, where matter distribution extends up to $R_b$, and beyond this radius, the spacetime is vacuum. The internal dynamics ($r<R_b$), where the spacetime is supported by an anisotropic fluid, play a crucial role in the complete perturbation analysis of the system. However, in this work our focus has been deliberately restricted to the exterior region ($r>R_b$) for the following reasons: (i) The interior JMN-1 spacetime ($r<R_b$) is already obtained as an equilibrium or stable configuration from gravitational collapse \cite{Joshi:2011zm}, (ii) Our primary aim is to investigate the stability of the spacetime under axial perturbations in the vacuum domain, where the perturbation equations take a tractable Regge–Wheeler type form. 
 Similarly, in the context of black holes, most of the literature (e.g., Schwarzschild, Kerr and Reissner–Nordström spacetime perturbations) has focused primarily on the exterior geometry, while the 
stability of the interior has largely been investigated using
quantum gravity methods \cite{Boehmer:2008fz}, (iii) The interior region involves non-vacuum anisotropic matter sources, which significantly complicates the perturbation analysis and requires a different methodology, typically involving coupled fluid–metric perturbations \cite{Andersson:2002jd,Gundlach:1999bt}. This separation of focus is a standard practice in perturbation theory, for example, in the case of the Sun and other stars, helioseismology studies often analyze oscillations by first treating the vacuum exterior separately \cite{ChristensenDalsgaard:2002, Kokkotas:1999bd}. Motivated by this well established tradition, we have here confined ourselves to the exterior geometry of compact objects (JMN-1 naked singularity). Furhtermore, within the matter-distributed region ($r < R_b$), ingoing waves are expected to be partially absorbed by the singularity for a certain range of the parameter $M_0$ \cite{Acharya:2023vlv}, or they may undergo partial reflection, producing echoes \cite{Chowdhury:2020}. We will address this issue in greater detail in a separate work. Moreover, the inner boundary may be considered down to scales of the order of the Planck length when analyzing the stability of the internal structure of this spacetime.

\section{WKB method} 
\label{sec:4}
\begin{figure}[h]
    \centering
    \includegraphics[width=0.45\textwidth]{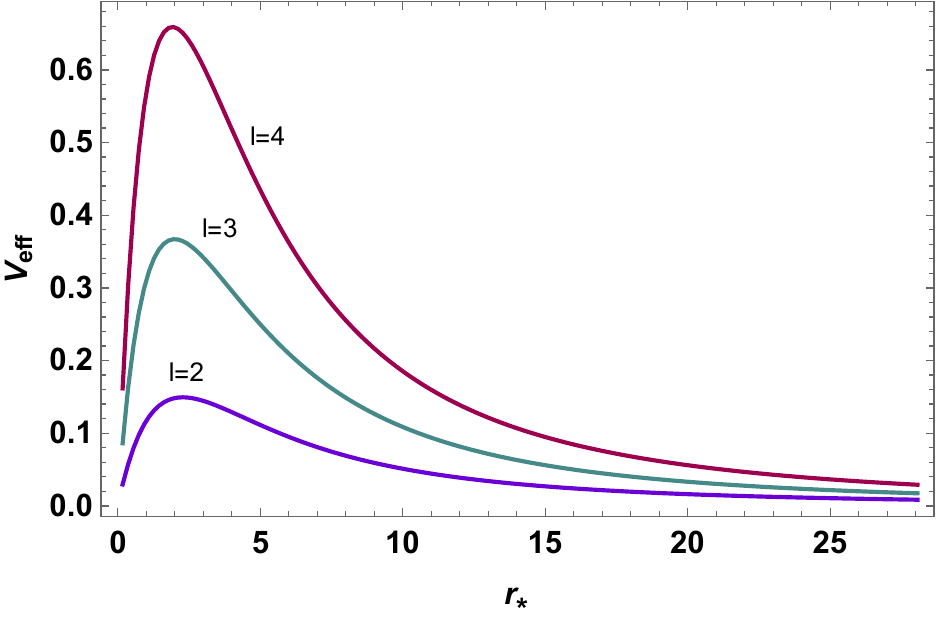}
    \caption{Parametric plot of the effective potential \( V_{\text{eff}}(r) \) as a function of the tortoise coordinate \( r_* \) for \( l = 2 \), \( l = 3 \), and \( l = 4 \) with \( M_0 = 0.7 \) and \( M = 1 \). The potential exhibits a barrier-like structure. Higher values of \( l \) lead to taller and narrower potential barriers.}
    \label{fig:effective_potential}
\end{figure}

To analyze the stability of the JMN-1 naked singularity under axial perturbations, we compute the quasi-normal mode frequencies using the Wentzel-Kramers-Brillouin (WKB) method. This method is effective for potentials that exhibit a barrier-like structure, as seen in our case, where the RW potential describes the perturbations around the singularity (see Fig.\ref{fig:effective_potential}). QNMs are the complex eigenmodes of the perturbations of spacetime, their real parts represent the oscillation frequencies, while the imaginary ones determine the damping rates of the modes.

The evolution of axial gravitational perturbations in the JMN-1 naked singularity background is governed by the master equation, 
\begin{equation}
\frac{d^2\psi}{dx^2} + Q(x)\psi = 0,
\end{equation}  
Here x can be identified as the tortoise coordinate ($r_*$)
and the function Q(x) is 
\begin{equation}
Q(x) = \omega^2 - V_{\text{eff}}(r_*).
\end{equation}  
The most general form of the effective potential  \( V_{\text{eff}}(r) \) under axial perturbations for the JMN-1 spacetime takes the form: 
\begin{equation}
{V_\text{eff}}(r) = f(r) \left[ \frac{l(l+1)}{r^2} + (1 - s^2) \frac{f'(r)}{r} \right].
\end{equation}

The parameter $s$ gives the spin weight of the perturbing field. Since we are dealing with  electromagnetic, and gravitational perturbations, we use $s =1$ and $s =2$, respectively. 
The behavior of $Q(x)$ dictates whether perturbations decay or persist, making its precise formulation essential in analysis of the stability of the singularity under the given perturbation.

The Schrödinger-like equation in curved spacetime was initially solved using the WKB approximation by Schutz and Will in 1985 \cite{Schutz_1985}. Later, in 1987, Iyer and Will \cite{Iyer_1987} extended the method to third order, improving its accuracy. In 2003, Konoplya \cite{Konoplya:2003} further refined the approach by extending it to the sixth order. More recently, in 2017, Matyjasek and Opala \cite{Matyjasek_2017} pushed the approximation up to the $13^{th}$ order. However, while higher-order approximations generally improve precision, researchers have noted that this is not always the case. In fact, in some cases using the $13^{th}$-order WKB approximation has been shown to produce results that differ significantly from those obtained with third- and sixth-order approximations. At higher orders, the solution diverges due to numerical instabilities, as errors significantly increase instead of converging 
 \cite{hatsuda2020}. 
Therefore, in this work we focus on $3^{rd}$, and $6^{th}$ WKB orders of approximation.

\begin{figure}[h]
    \centering
    \includegraphics[width=0.48\textwidth]{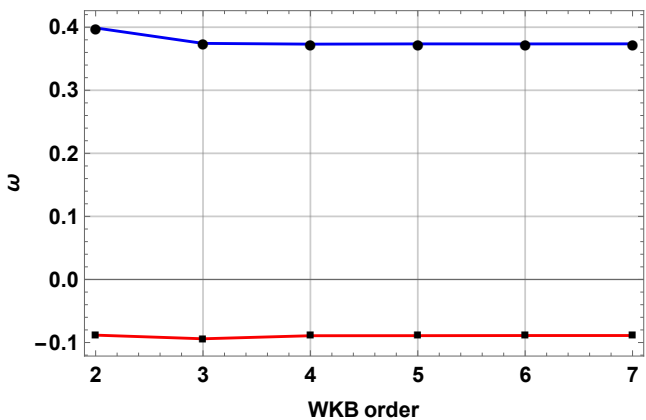}
    \caption{Convergence of quasinormal mode frequencies with increasing WKB order for obertone number $n=0$, $l =2$ amd $s =2$. The red color corresponds to imaginary and the blue color corresponds to real QMN frequencies.}
    \label{fig: 2}
\end{figure}

The third-order formula for the WKB approximation method is given by  
\begin{equation}
    \omega^2 = \left[ V_0 + \sqrt{-2V''_0} \right] - i \left(n+\frac{1}{2} \right) \sqrt{-2V''_0} (1+\Omega),
\end{equation}
where  

\begin{equation}
\Lambda(n) = \frac{1}{\sqrt{2Q''_0}} 
\left[ \frac{Q^{(4)}_0}{8Q''_0} 
\left( \frac{1}{4} + \alpha^2 \right) 
- \frac{1}{288} \frac{(Q''_0)^2}{(Q'_0)^2} 
(7 + 60\alpha^2) \right]
\label{eq:Lambda}
\end{equation} 

\begin{align}
    \Omega(n) &= \frac{(n+1)}{2} 
    \left( \frac{Q^{(4)}_0}{2 Q''_0} \right) 
    \left(\frac{51 + 100\alpha^2}{5} \right) \notag \\
    &\quad + \frac{5}{6912} \frac{(Q'''_0)^4}{(Q''_0)^4} (77 + 188\alpha^2)  
    - \frac{1}{384} \frac{(Q'''_0)^2}{(Q''_0)^3} (67 + 68\alpha^2) \notag \\
    &\quad + \frac{1}{288} \frac{Q'''_0 Q^{(5)}_0}{(Q''_0)^2} (19 + 28\alpha^2) 
    - \frac{1}{288} \frac{Q^{(6)}_0}{(Q''_0)^2} (5 + 4\alpha^2).
\end{align}
The sixth-order formula for the WKB approximation method is given by  
\begin{equation}
    i \frac{\omega^2 - V_0}{\sqrt{-2V''_0}} = \sum_{i=2}^{6} \Lambda_i,
\end{equation}
where,  
\begin{equation}
    \Lambda_i = \left(n+\frac{1}{2}\right), \quad n = 0,1,2, \dots.
\end{equation}
The $7^{th}$ and $8^{th}$ order WKB approximations represent a significant refinement in the semi-analytic computation of quasinormal modes, extending beyond the well-established sixth-order corrections. These higher-order terms, denoted as \(\Lambda_7\) and \(\Lambda_8\), incorporate additional derivatives of the potential function \( Q(x) \) at its peak, further improving the accuracy of frequency predictions. However, this extension comes at the cost of substantial computational complexity—the number of terms in \(\Lambda_7\) reaches 616, while \(\Lambda_8\) contains 1,215 terms. The general formulation remains consistent with previous orders,  
\[
i \frac{Q_0}{\sqrt{2Q''_0}} - \sum_{k=2}^{N} \Lambda_k = n + \frac{1}{2},
\] 
where each \(\Lambda_k\) contributes increasingly precise corrections to the fundamental frequency. The accuracy of these approximations has been validated through direct comparisons with numerical results, demonstrating significant improvements over the sixth-order formulation. While the algebraic complexity of \(\Lambda_7\) and \(\Lambda_8\) makes direct manipulation cumbersome, their inclusion leads to a noticeable reduction in deviations from full numerical solutions (see Figure \ref{fig: 2}).

The presence of increasingly large terms in the expansion can lead to numerical instabilities, as observed in the $12^{th}$-order results. 
This highlights the trade-off between precision and practicality when using the WKB method. In our analysis, we find that the sixth and seventh orders provide the most reliable results before numerical errors start dominating. The convergence as also seen in Figure \ref{fig: 2} as these orders aligns well with previous studies on WKB-based QNM computations \cite{Konoplya202X}, reinforcing their applicability in gravitational wave research. Moreover, the deviation at extremely high orders suggests that alternative resummation techniques, such as Padé approximants, could be explored to enhance the stability of WKB calculations further. These refinements may help in extending the method’s validity to more extreme astrophysical scenarios, including highly compact objects and beyond-GR modifications. In this paper, we calculate the QNMs using the numerical WKB package \cite{Konoplya202X}.

\section{The QNMs frequencies}
\label{sec:resullts}
In this section, we present the results of our numerical calculations. 
Tables \ref{tab1QNMs} to \ref{tab:QNM_WKB} display the real and imaginary parts of the QNM frequencies for the two primary types of axial perturbations: electromagnetic and gravitational. Electromagnetic perturbations $(s = 1)$ correspond to the perturbations of the Maxwell field in the given background. 
Finally, gravitational perturbations $(s = 2)$ are associated with fluctuations in the metric tensor itself, representing the fundamental response of the spacetime geometry to small disturbances. The data from the tables show imaginary negative parts of the QNM frequencies across all cases, therefore, confirming the stability of the spacetime.

\begin{table}[h]
    
    \small  
    \renewcommand{\arraystretch}{1.1}  
    \setlength{\tabcolsep}{4pt}  
    \caption{Real and imaginary parts of the fundamental QNM frequencies in gravitational perturbations ($s=2$) for $3^{rd}$ order WKB approximation.}
    \begin{tabular}{|c|c|c|c}
        \toprule
        $M_0$ & $l=2, n=0$ & $l=3, n=1$ \\
        
        & $\omega (WKB)$ & $\omega (WKB)$ \\
        
        \hline
        0.1 & $0.373162 -0.0892174 i$ & $0.582355 -0.281406 i$ \\
    
        0.3  & $0.373199 -0.0892264 i$ & $0.582413 -0.281434   i$ \\
        
        0.5 & $0.373162 -0.0892174  i$ & $0.582355 -0.281406 i$ \\
        0.6  & $0.373199 -0.0892264 i$ & $0.582413 -0.281434 i$ \\
        0.7 & $0.370831 -0.0886602 i$ & $0.582363 -0.28141  i$ \\
       \hline
    
    \end{tabular}
    \label{tab1QNMs}
\end{table}

\begin{table}[h]
    
    \small  
    \renewcommand{\arraystretch}{1.1}  
    \setlength{\tabcolsep}{4pt}  
    \caption{Real and imaginary parts of the fundamental QNM frequencies in gravitational perturbations ($s=2$) for $6^{th}$ order WKB approximation.}
    \begin{tabular}{|c|c|c|c}
        \toprule
        $M_0$ & $l=2, n=0$ & $l=3, n=1$ \\
        
         & $\omega (WKB)$ & $\omega (WKB)$ \\
        
         \hline
        0.1  & $0.373619 -0.088891 i$ & $0.582642 -0.28129 i$ \\
        0.3  & $0.373657 -0.0888999 i$ & $0.5827 -0.281319    i$ \\
        0.5  & $0.373619 -0.088891i$ & $0.582642 -0.28129 i$ \\
        0.6  & $0.373657 -0.0888999i$ & $0.5827 -0.281319  i$ \\
        0.7  & $0.371286 -0.0883358 i$ & $0.579003 -0.279534    i$ \\
        \hline
        
    \end{tabular}
    \label{tab2QNMs}
\end{table}

\begin{table}[h]
    
    \small  
    \renewcommand{\arraystretch}{1.1}  
    \setlength{\tabcolsep}{4pt}  
    \caption{Real and imaginary parts of the fundamental QNM frequencies in electromagnetic perturbations ($s=1$) for $6^{th}$ order WKB approximation.}
    \begin{tabular}{|c|c|c|c}
        \toprule
        $M_0$ & $l=2, n=0$ & $l=3, n=1$ \\
        
         & $\omega (WKB)$ & $\omega (WKB)$ \\
        
         \hline
        0.1  & $0.457593 -0.0950112 i$ & $0.641737 -0.289731 i$ \\
        0.3  & $0.457639 -0.0950207  i$ & $0.641801 -0.28976     i$ \\
        0.5  & $0.457593 -0.0950112i$ & $0.641737 -0.289731i$ \\
        0.6  & $0.457639 -0.0950207i$ & $0.641801 -0.28976  i$ \\
        0.7  & $0.454735 -0.0944177   i$ & $0.637728 -0.287921    i$ \\
        \hline
        
    \end{tabular}
    \label{tab4QNMs}
\end{table}

\begin{table}[h]
    \centering
    \renewcommand{\arraystretch}{1.2}
     \caption{Quasinormal mode frequencies for $M_0 = 0.7$, $n=0$, $l =2$ and $s=2$ and error estimations for different WKB orders.}
     \vspace{1mm}
    \begin{tabular}{|c|c|c|}
        \hline
        \textbf{WKB order} & \textbf{QNMs} & \textbf{Error} \\
        \hline
        
        9  & 0.37166 - 0.088915$i$ & 0.0958424 \\
        8  & 0.371489 - 0.088917$i$ & 0.00386661 \\
        7  & 0.37126 - 0.0882506$i$ & 0.000123047 \\
        6  & 0.371286 - 0.0883358$i$ & 0.0000732704 \\
        5  & 0.371172 - 0.088363$i$ & 0.000120327 \\
        4  & 0.37122 - 0.0885673$i$ & 0.000225863 \\
        3  & 0.370831 - 0.0886602$i$ & 0.00248593 \\
        \hline
    \end{tabular}
    \label{tab:QNM_WKB}
\end{table}

For gravitational (s=2), and electromagnetic (s=1) perturbations, the real parts of the QNMs remain nearly constant, with a slight decrease at $M_0=0.7$ as shown in tables \ref{tab1QNMs},\ref{tab2QNMs}, \ref{tab4QNMs}, and \ref{tab:QNM_WKB}. The imaginary parts also show minimal variation, this indicates a stable decay rate across all cases. These trends suggest that the JMN1 spacetime maintains its stability without significant deviations in the QNM spectrum.

Note that table \ref{tab:QNM_WKB} presents quasinormal mode (QNM) frequencies computed using the WKB approximation from the $3^{rd}$ to the $6^{th}$ order. The real part represents oscillation frequency, while the imaginary part determines damping. As stressed in Section \ref{sec:4}, higher-order WKB approximations generally improve accuracy, as seen from decreasing error estimates. Lower-order approximations ($3^{rd}$–$5^{th}$) show lower fluctuations and errors, whereas the $7^{th}$ and $8^{th}$ orders exhibit convergence, making them the most stable and reliable. The $12^{th}$-order deviates significantly, likely due to limitations of asymptotic expansion at high orders. Thus, the $7^{th}$–$8^{th}$ order WKB approximation provides the best balance between accuracy and reliability for computing QNMs.

 
\begin{figure}[h]
    \centering
    \hspace{-0.9cm}
    \label{collective}
\end{figure}

 When a system undergoes an initial perturbation, its response carries crucial information about its stability and underlying structure. In the analysis of the JMN-1 naked singularity, axial gravitational perturbations evolve over time, exhibiting a characteristic decay pattern. Although the time-dependent evolution of the perturbation is not explicitly depicted here, it typically reveals some important dynamical features. Initially, the perturbations undergo a phase of quasinormal ringing, where oscillations gradually decrease in amplitude. This stage is followed by a decay phase, where the perturbations diminish at a slower rate. The rate at which perturbations dissipate depends strongly on the angular momentum mode $l$. 

\begin{figure}[h]
    \centering
    \label{Schwarzschild}
\end{figure}


 \section{Discussions and Conclusions}
\label{sec:5}

After the discovery of gravitational waves, the study of compact objects has gained immense significance, leading to extensive investigations into various alternatives to black holes. While the event horizon remains a defining feature of classical black holes, numerous theoretical models propose horizonless ultracompact objects. These objects could mimic black hole behavior while exhibiting distinct physical signatures. One such alternative is the JMN-1 naked singularity, which arises from the gravitational collapse of a massive star under specific conditions. Investigating the stability and observational imprints of such spacetime is thus of crucial importance in modern gravitational physics. In this paper, we have considered JMN-1 naked singularity to check whether it is stable under small perturbations or not.

To assess the stability of this naked singularity, we studied its behavior under linear axial perturbations such as electromagnetic, and gravitational. Then we have extracted the QNMs frequencies, which are the eigenmodes of the perturbations. By employing the methods of classical perturbation theory, we expanded the metric and retained only the linear terms to obtain the master equation. We impose appropriate boundary conditions as discussed in Sec.(\ref{sec:3}), namely an inner boundary condition at $r \to R_b$  and an outer boundary condition at $r \to 100 M$, which is comparable to spatial infinity \cite{Kokkotas:1999bd, Berti:2009kk}.

We would like to point out that the present analysis has been conducted for the exterior region of the JMN-1 spacetime. Specifically, we have focused on the vacuum solution extending from the matching radius up to spatial infinity as discussed above. This approach is consistent with existing literature on the computation of QNM frequencies of black holes, where the analysis is typically restricted to the exterior geometry, from the event horizon (e.g., \( r = 2M \) for Schwarzschild) to spatial infinity. The rationale behind this is that observational signatures, such as QNM frequencies, originate from perturbations in the exterior region, making this domain particularly relevant for gravitational wave detection. However, it is essential to note that investigating interior stability is equally important. We intend to address this in future work by incorporating quantum gravity-inspired models. In particular, we plan to explore effective geometries such as the Nariai-type universe or modifications arising from loop quantum gravity, which may provide a meaningful framework for analyzing the stability of the interior region and addressing the central singularity.

The potential of JMN-1 naked singularity is well-behaved and features a barrier structure that temporarily traps the incoming and outgoing waves. In addition, the function $Q(x, \omega) $ has two turning points. Therefore, we applied the WKB method to solve the master equation and compute the QNM frequencies $\omega$. These frequencies are complex in nature: the real part corresponds to the oscillation frequency, while the imaginary part determines the damping rate. Over the years, the WKB method has been extensively developed to solve such differential equations in the context of gravitational perturbation theory. We found that our results are best captured by orders between 3rd and 6th, whereas deviations occur at higher order approximations.  We 
observe that our WKB approach at these orders
demonstrates strong convergence when we consider higher orders such as $7^{th}$ and $8^{th}$
(see Fig. \ref{fig: 2}). 
This clearly suggests that the WKB method is highly effective within a certain range of approximations. Moreover, we shoud note that in some cases much higher-order terms can lead to numerical instabilities or yield diminishing improvements in accuracy \cite{hatsuda2020, Zhao:2024}.

It is worth noting that we did not compute the QNM frequencies for the fundamental mode $l = n = 0$, as the WKB method is only valid when $l > n$ and does not yield precise results for the fundamental mode. Nevertheless, our results demonstrate that the imaginary part of the QNM frequencies remains negative, indicating that the JMN-1 naked singularity is stable under linear perturbations. Furthermore, our numerical simulations confirm the stability of the spacetime under electromagnetic $(s=1)$, and gravitational $(s=2)$ perturbations.
This is confirmed from tables \ref{tab1QNMs},\ref{tab2QNMs}, \ref{tab4QNMs}, and \ref{tab:QNM_WKB} where the imaginary part of the QNM frequency $\omega$ remains negative, indicating damping rather than exponential growth.  

Note that we have 
explored the parameter regime $4/5>M_0 > 2/3$ and $R_b \leq 3M$ of the JMN-1 naked singularity, as this is where a photon sphere is present and shadow formation occurs \cite{Saurabh:2023otl}. This analysis is crucial for ensuring the stability of the JMN-1 naked singularity in the presence of a photon sphere and is also significant in the context of future EHT observations \cite{EventHorizonTelescope:2022xqj}.

In summary, we conclude that the JMN-1 naked singularity is stable under axial linear perturbations. Future research could focus on investigating the implications of this stability in the context of observational constraints and gravitational wave signatures. It is important to emphasize that our present analysis has focused exclusively on the axial (odd-parity) sector, following the standard approach in perturbation theory where axial stability is often treated as the first step \cite{Regge:1957td, Moncrief:1974ng, Noui:2023ksf}. A comprehensive stability assessment also requires the study of polar (even-parity) perturbations, which we plan to address in forthcoming work. Together, these analyses will provide a complete picture of the stability properties of the spacetime.

With the advent of these next-generation gravitational wave detectors, precise measurements of QNMs from ultracompact objects, including naked singularities, could shed light on the nature of strong field gravity and potential deviations from classical black hole physics. In addition to that, exploring the role of quantum gravity effects in shaping the near singularity dynamics could provide deeper insights into the fundamental nature of spacetime. 

\acknowledgments{P. Bambhaniya and Elisabete M. de Gouveia Dal Pino acknowledge support from the São Paulo State Funding Agency FAPESP (grant 2024/09383-4). Elisabete M. de Gouveia Dal Pino also acknowledges the support from the São Paulo State Funding Agency FAPESP (grant 2021/02120-0) and CNPq (grant 308643/2017-8). A. Pathrikar expresses his gratitude to ICSC, Ahmedabad University, for giving an opportunity to visit and facilitating discussions on this project. P. Bambhaniya would like to thank Anurag Patel, Kimet Jusufi, Dipanjan Dey and Saurabh for discussions and their valuable suggestions on this work.}

\end{document}